\documentclass[traditabstract]{aa} 
\usepackage{lscape}
\usepackage{graphicx}
\usepackage{txfonts}
\usepackage{natbib}
\bibpunct{(}{)}{;}{a}{}{,}

\newcommand\fho{\mbox{$^{\mathrm h}$}}%
\newcommand\fmi{\mbox{$^{\mathrm m}$}}%
\newcommand{\cmmth}{\mbox{cm$^{-3}$}}
\newcommand{\cmmtw}{\mbox{cm$^{-2}$}}

\newcommand{\kms}{\mbox{km\,s$^{-1}$}}
\newcommand{\msun}{\mbox{$M_{\odot}$}}
\newcommand{\lsun}{\mbox{$L_{\odot}$}}

\begin{document}
\titlerunning{Observations of methanol in the IRDC core G11.11-0.12P1}
\authorrunning{L. G\'omez et al.}

   \title{High-angular resolution observations of methanol in the 
     infrared dark cloud core G11.11-0.12P1\thanks{Based on observations 
       carried out with the IRAM Plateau de Bure Interferometer and the 
       IRAM~30m telescope. IRAM is supported by
    INSU/CNRS (France), MPG (Germany), and IGN (Spain).} }

   \author{Laura G\'omez
          \inst{1}\fnmsep\thanks{Member of the International Max Planck
          Research School (IMPRS) for Astronomy and Astrophysics at
	  the Universities of Bonn and Cologne.}
          \and
          Friedrich Wyrowski\inst{1}
	  \and
         Thushara Pillai\inst{2}
         \and
         Silvia Leurini\inst{1}
          \and 
          Karl M. Menten\inst{1}
          }

   \institute{Max-Planck-Institut f\"ur Radioastronomie, Auf dem H\"ugel 69,
              D-53121 Bonn, Germany\\
              \email{[lgomez, wyrowski, sleurini, kmenten]@mpifr.de}
                        \and
             California Institute of Technology, 
             MC 249-17, 1200 East California Boulevard, Pasadena, CA 91125,
             USA\\
             \email{tpillai@astro.caltech.edu}
            }

   \date{Received ; accepted }

  \abstract
   {Recent studies suggest that infrared dark clouds (IRDCs) have 
    the potential of harboring
    the earliest stages of massive star formation and indeed evidence for
    this is found toward distinct regions within them.
   We present a study with the Plateau de Bure Interferometer 
   of a core in the archetypal filamentary IRDC G11.11-0.12
   at few arcsecond resolution to determine its physical and chemical structure. 
   The data consist of continuum and line observations
   covering the C$^{34}$S 2 $\rightarrow$ 1 line and
   the methanol 2$_{k} \rightarrow$ 1$_{k}~v_{t} = 0$ lines at 3mm and
   the methanol 5$_{k} \rightarrow$ 4$_{k}~v_{t} = 0$ lines at 1mm.  
   Our observations show extended emission in the continuum at 1 and 3 mm. 
   The methanol 2$_{k} \rightarrow$ 1$_{k}$  $v_{t} = 0$ 
   emission presents three maxima extending over 1 pc scale
   (when merged with single-dish short-spacing observations); 
   one of the maxima is spatially coincident with the continuum emission.
     The fitting results show 
  enhanced methanol fractional abundance ($\sim$3~$\times$~10$^{-8}$) at 
  the central peak with respect to the other two peaks, where it
  decreases by about an order of magnitude ($\sim$4-6~$\times$~10$^{-9}$).
     Evidence of  extended 4.5 $\mu$m emission, ``wings'' in
        the CH$_3$OH 2$_{k} \rightarrow$ 1$_{k}$ spectra, and CH$_3$OH abundance enhancement point
      to the presence of an outflow in the East-West direction.
   In addition, we find a gradient of $\sim$4 \kms~in the same direction,
  which we interpret as being produced by an outflow(s)-cloud interaction.
}


   \maketitle
%

\section{Introduction}
    The objects now termed Infrared Dark Clouds (IRDCs) were first identified 
    by \citet{pe96} and \citet{eg98} in images of 
    the Galactic plane made with the {\it Infrared Space Observatory} (ISO) and 
    the {\it Midcourse Space Experiment} (MSX) satellite, respectively. IRDCs 
    have significant extinctions even at 8 and 15 $\mu$m and are, thus, seen
    in silhouette against the bright, diffuse, mid-infrared Galactic emission.
    Millimeter and (sub)millimeter molecular studies reveal that some
    clumps within IRDCs have
    column densities of $>$ 10$^{23}$ \cmmtw, volume gas
    densities of $n >$ 10$^{5}$ \cmmth, and temperatures $T <$ 20 K  
    \citep[e.g.,][]{ca98,pi06a}. Several studies suggest that IRDCs have 
    the potential 
    of harboring the earliest stages of massive star formation and indeed 
    evidence for this is found toward distinct regions within them  
    \citep{ra05,pi06b,wa06}.    
    
\begin{figure*}
   \centering
   \includegraphics[angle=-90,width=17.5cm]{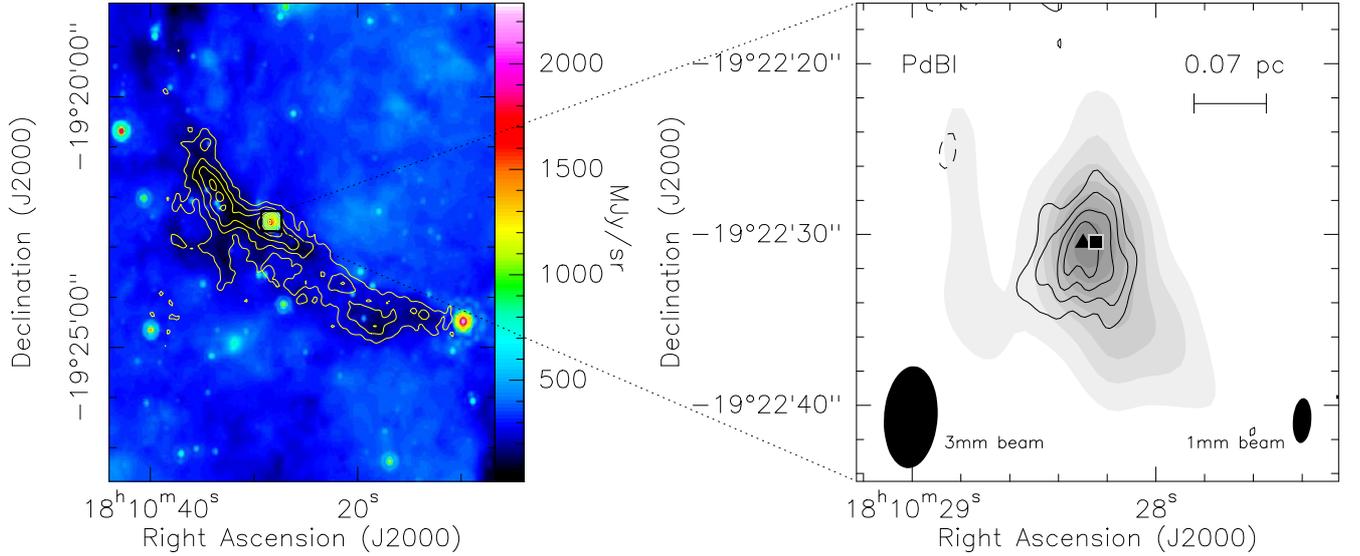}
      \caption{
        {\it Left}: SCUBA 850 $\mu$m \citep{ca00} contours overlaid on the 
        MIPSGAL 24 $\mu$m 
        image of the IRDC G11.11-0.12. {\it Right}:
        PdBI 1.2 mm continuum emission (contours) of G11.11P1 overlaid
        on the 3.1 mm  continuum emission (grey scale image). 
        First contour and contour spacing for the 3.1 mm emission are 0.9 mJy 
        beam$^{-1}$ (3$\sigma$), the dotted contours show the negative emission
        (-3$\sigma$); the synthesized beam 
        (6\farcs0 $\times$ 3\farcs1; PA = 
        176\degr) is shown in the bottom left corner. 
        First contour and contour spacing for the 1.2 mm emission are 6.3 
        mJy beam$^{-1}$ (3$\sigma$), the dashed contours show the negative 
        emission (-3$\sigma$); the synthesized beam 
        (2\farcs6 $\times$ 1\farcs1; PA =
        176\degr) is shown in the bottom right corner. The filled triangle
        at R.A. = 18\fho10\fmi28\fs29, Decl. = 
        -19\degr22\arcmin30\farcs5 and the filled square at
        R.A. = 18\fho10\fmi28\fs25, Decl. =
        -19\degr22\arcmin30\farcs45 (J2000) indicate the water and 
        (integrated emission) methanol masers, 
        respectively, reported by Pillai et al. (2006b).
              }
         \label{Fcont}
   \end{figure*}

    \citet{pi06a} mapped the ($J,K$) $=$ (1,1) and (2,2) inversion transitions
    of ammonia (NH$_3$) in the IRDC G11.11-0.12 with the Effelsberg 100m 
    telescope
    and found gas temperatures of the order of 15 K for several clumps
    within this cloud. The {\it left} panel of Fig.~\ref{Fcont} shows an 
    overview of the region. Furthermore, \citet{pi06b} reported the detection 
    of 6.7 GHz class II methanol (CH$_3$OH) and 22.2 GHz water (H$_2$O) masers 
    in the IRDC Core G11.11-0.12P1 (hereafter G11.11P1). Both CH$_3$OH and 
    H$_2$O masers are known as tracers of massive star formation. They found 
    that these masers were associated with a SCUBA dust continuum peak 
    \citep[P1;][]{ca00} and ascribed the kinematics of the methanol masing 
    spots to a maser amplification from a Keplerian disk.

    Recently, \citet{he10} observed this filamentary IRDC with the PACS 
    (at 70, 100, and 160 $\mu$m) and the SPIRE (250, 350, and 500 $\mu$m) 
    instruments on board of the {\it Herschel Space Observatory},
    with resolutions in the range of $\sim$6\arcsec--40\arcsec. For G11.11P1,
    they obtain from a modified blackbody fit to the PACS data, a dust temperature of 
    24 K, a luminosity of 1346 \lsun, and a mass of 240 \msun.  
    \citet{pi06a} and \citet{le07b} have used a two-component model to fit
    single-dish observations of NH$_3$ and CH$_3$OH spectra, 
    respectively, and derived temperatures of $\sim$15--18 K for the 
    extended component (with size of $\sim$20\arcsec~corresponding to
    0.35~pc) and
    $\sim$47--60 K for the inner component (with a size of
    $\sim$3\arcsec~corresponding to 0.05~pc).
    
    Methanol, which is a slightly asymmetric top molecule, has been
    used to derive physical parameters such as density and temperature
    in IRDCs \citep{le07b}, high-mass protostellar objects 
    \citep[e.g.,][]{le07a}, massive young 
    stars \citep[e.g.,][]{vdt00} as well as in low-mass protostellar systems 
    \citep{kr10}. In particular, \citet{vdt00} carried out a study toward 
    13 massive 
    star-forming regions and found three types of CH$_3$OH abundance 
    ($X_{\rm CH_3OH} \equiv N_{\rm CH_3OH}/N_{\rm{H_2}}$) profiles: 
    $X_{\rm CH_3OH} \sim 10^{-9}$ for the coldest sources, 
    from $10^{-9}$ to $10^{-7}$ for warmer sources, and $10^{-7}$ for hot cores.
    CH$_3$OH has also been associated with outflows where 
    the methanol abundance enhancements have been found to be a factor as large
    as 400-1\,000 \citep[e.g.,][]{ba95,ga02,kr10}. 

    In this paper, we present arcsecond resolution millimeter continuum and 
    line observations toward G11.11P1 with the aim to determining its physical 
    and chemical structure. In Sect.~2, we describe our observations
    carried out with the IRAM Plateau de Bure Interferometer. In Sect.~3, we 
    present continuum and line results together with the analysis. The 
    discussion is presented in Sect.~4 and the summary is given in Sect.~5.

\section{Observations and data reduction}

\subsection{IRAM PdBI observations}
    G11.11P1 was observed with the IRAM
    six element array Plateau de Bure 
    Interferometer \citep[PdBI;][]{gu92} in D
    and C configurations in 2005 and 2006, respectively. 

    During the 2006 observations,
    one antenna was equipped with the prototype New Generation Receiver.
    Because of a difference in the frequency scheme of this receiver,
    visibilities obtained in the image sideband in this antenna were discarded.
    The receivers were tuned single side-band at 3 mm and double side-band at
    1 mm. The 3 mm receivers were centered at 96.64 GHz  
    and the 1 mm receivers at 241.81 GHz (see Table~\ref{Tobs}). At 3 mm, 
    the C$^{34}$S 2 $\rightarrow$ 1 line and
    the CH$_{3}$OH 2$_{k} \rightarrow$ 1$_{k}$ $v_{t} = 0$ lines were
    covered by using two correlator units of 80 MHz with a spectral
    resolution of 0.3125~MHz (0.97 \kms) and by two correlator units
    of 320 MHz. The latter were used, free of line, to produce
    the continuum image with a total bandwith of $\sim$450~MHz. 
    At 1 mm, the CH$_{3}$OH 5$_{k} \rightarrow$ 4$_{k}$ $v_{t} = 0$   
    lines were observed with two units of 160 MHz with a spectral 
    resolution of 1.250 MHz (1.55 \kms). The remaining units of 80, 160, 
    and 320~MHz were placed in such a way that a frequency range free of lines 
    could be used to measure the continuum flux. The total bandwidth of both
    sidebands was $\sim$700~MHz.
    
    The phase and amplitude were calibrated with observations of the
    object 1730-130. 
    The bandpass calibration was done with 3C~273. MWC~349 was used as 
    primary flux calibrator of the 3 and 1 mm data (see Table~\ref{Tobs}).
        We estimate the final flux density accuracy to be $\sim$5\% and 10\%
    for the 3mm and 1mm data, respectively.
    Continuum images 
    were subtracted from the line data in the visibility plane. 
    The combination of the C and D configurations provides
    angular scales from 1\farcs8--12\farcs6 (1 mm) and 
    4\farcs4--32\farcs5~(3 mm), i.e., providing information on spatial
    scales of 0.03--0.22 pc and 0.08--0.57 pc, respectively.
       
    The calibration and data reduction were made within the GILDAS 
    package\footnote{http://www.iram.fr/IRAMFR/GILDAS} at IRAM Grenoble. 
    Images were created with natural weighting and CLEANed using the
    standard H\"ogbom algorithm.

\begin{table}
\caption{Parameters for the IRAM PdBI observations.}
\label{Tobs}
\centering
\begin{tabular}{l c} 
\hline\hline
\noalign{\smallskip}
Parameter & Value\\
\noalign{\smallskip}
\hline 
\noalign{\smallskip}
Pointing center    &  R.A. (J2000) = 18\fho10\fmi28\fs24\\
                   &  Decl. (J2000) = -19\degr22\arcmin30\farcs5\\
Number of antennas &  6\\
Baseline range    &  24-176 m\\
Band center        &   96.64  and 241.81 GHz\\
Primary HPBW       &  52\arcsec~at 96.64 GHz\\
                   &  21\arcsec~at 241.81 GHz\\
Synthesised HPBW   &  $6\farcs0 \times 3\farcs1$ at 96.64 GHz\\ 
                   &  $2\farcs6 \times 1\farcs1$ at 241.81 GHz\\
Primary flux density calibrator: &  1.17 Jy at 96.64 GHz\\
MWC349                           &  2.03 Jy at 241.81 GHz\\
\noalign{\smallskip}
\hline                   
\end{tabular}
\end{table}

\begin{table*}
\begin{minipage}{\textwidth}
\caption{Parameters of the observed continuum emission.}
\label{Tgauss}
\centering
\renewcommand{\footnoterule}{} 
\begin{tabular}{rcccccc}
\hline \hline
\noalign{\smallskip}
Frequency &  \multicolumn{2}{c}{Peak Intensity Position}  & Peak Intensity & Flux density\footnote{Flux density was obtained by summing the continuum emission above 3$\sigma$ level within a polygon encompassing the core.}  & Deconvolved angular size\footnote{Major axis $\times$ minor axis, at FWHM; position angle of major axis from fits of elliptical Gaussian.} &Mass\footnote{Assuming a distance of 3.6 kpc and dust temperature of 60 K.} 
\\
(GHz)     & R.A. (J2000) & Decl. (J2000)  & (mJy beam$^{-1}$)  &(mJy)  & &($M_{\odot}$)\\
\noalign{\smallskip}
\hline
\noalign{\smallskip}
   96.7  & 18\fho10\fmi28\fs28 & $-$19\degr22\arcmin30\farcs5 & \phantom{0}7.0 & \phantom{0}20.2  & $(8\farcs9\pm0\farcs6)\times (4\farcs4\pm0\farcs5);~+24^\circ\pm 7^\circ$  &  37\\     
   241.8 & 18\fho10\fmi28\fs29& $-$19\degr22\arcmin30\farcs5 & 30.8 & 208.6  & $(4\farcs9\pm0\farcs6)\times (3\farcs9\pm0\farcs5);~+165^\circ\pm 36^\circ$  & 13\\
\noalign{\smallskip}
\hline
\end{tabular}
\end{minipage}
\end{table*}

\subsection{IRAM 30m short-spacing observations}\label{g11:sp}
    G11.11P1 was observed  
    in the 2$_{k}\rightarrow$1$_{k}$ $v_{t} = 0$ (3 mm) and
    5$_{k} \rightarrow$ 4$_{k}$ (1 mm) CH$_{3}$OH bands
    with the IRAM 30m telescope. An area of 90\arcsec~$\times$ 90\arcsec 
    was mapped with the SIS B100 receiver at 3 mm and 
    66\arcsec~$\times$ 66\arcsec~with the HERA receiver \citep{su04} at 1 mm
    with a bandwidth of $\sim$140~MHz. Sampling at 3 mm was of 
    15\arcsec~while at 1 mm was of 6\arcsec. Conversion from antenna 
    temperature to main-beam brightness temperature ($T_{\rm MB}$) was 
    performed by using a beam efficiency of 0.78 at 3 mm and 0.52 at 1 mm. 
    The $rms$ at 1 mm is 0.3 K and at 3 mm is 0.02 K. The data were reduced 
    with the CLASS program, part of the GILDAS software package.

    The 3 mm single-dish data are used to create short-spacing
    pseudo-visibilities; these are then merged with the interferometric 
      observations. All data are imaged and deconvolved together.
    The processing was performed in GILDAS/MAPPING following standard
    procedures of the UV\_SHORT task (see Rodr\'iguez-Fern\'andez et al. 
    2008\footnote{http://iram.fr/GENERAL/reports/IRAM\_memo\_2008-2-short-spacings.pdf} for details on the pseudo-visitbility technique).
    In Sect.~\ref{smerged} we present the result
    of recovering the short-spacing information for the 3 mm data.

    Unfortunately, we were not able to merge the interferometric and
    single-dish observations for the 1 mm data due to poor 
    signal-to-noise ratio of the latter data. 
    However, we make use of the 1 mm single-dish data in the following 
    sections to help in the analysis and interpretation of this core.

\section{Results and Analysis}
\subsection{PdBI: Continuum emission}

\subsubsection{Morphology and core size} 

    The {\it right} panel of Fig.~\ref{Fcont} presents the continuum 
    images at 1 and 3 mm of G11.11P1. At the resolution of both images, 
    the emission is extended; 
    the peak intensity positions coincide with each other
    and with the maser positions as well. 
    The morphology at 3 mm resembles that previously imaged with the 
    Berkeley-Illinois-Maryland-Association (BIMA) interferometer by 
    \citet{pi06b} with a slightly lower resolution 
    ($8\farcs3 \times 3\farcs9$). 
    
    In Table~\ref{Tgauss} we list the parameters of the continuum emission.
    Deconvolved sizes have been obtained by fitting 
    two-dimensional Gaussians to the continuum maps and yielding a 
    core size of $\sim$0.16 pc $\times$ 0.08 pc with a PA $24^\circ$ at 
    3 mm and $\sim$0.09 pc $\times$ 0.07 pc with a PA $165^\circ$ at 1 mm.
    We have assumed a kinematic distance of 3.6 kpc.

\subsubsection{Mass}

    Assuming that the mm continuum is mainly due to 
    optically thin dust emission, we estimate the 
    gas mass, $M_g$, by using the following expression \citep{hi83}
    \begin{equation}
      M_g = \frac{F_{\nu}~D^2~R_{\rm gd}}{\kappa_{\nu}~B_{\nu}(T_{\rm d})} \label{emass}\,~\msun,
   \end{equation}
    where $F_{\nu}$ is the observed integrated flux density in Jy, $D$ is the
    distance, $R_{\rm gd}$ is the gas-to-dust mass ratio, $\kappa_{\nu}$ 
    is the dust opacity coefficient, and $B_{\nu}$($T_{\rm d}$) is the 
    Planck function at the dust temperature ($T_{\rm d}$). The 
    1 mm flux density was obtained by summing the 1 mm continuum 
    emission above 
     3$\sigma$ ($>$ 6 mJy beam$^{-1}$) level within a polygon encompassing 
    the core; the integrated flux density is $\sim$209 mJy. We assume a 
    gas-to-dust mass ratio of 100, and adopt a $\kappa_{\nu} =$ 1
    cm$^2$~g$^{-1}$ \citep{os94}, at 1mm, for an MRN \citep{mnr}
    graphite-silicate grain mixture with thick ice mantles, 
    at a gas density of 10$^{6}$ \cmmth. 
    If we use $T_{\rm d} =$ 60 K \citep{le07b,pi06a}, a proper temperature 
    for a region where methanol emission is arising, a mass of 13 \msun~at 1 mm
    is obtained for G11.11P1 (see Table~\ref{Tgauss}). 
    This is equivalent to adopt a dust opacity coefficient 
    $k_\nu = 0.1~(\nu/10^{12}~{\rm Hz})^\beta$ \citep{bec90}, which includes
    the gas contribution to the total mass for a $R_{\rm gd}=100$, 
    with $\beta =$ 1.6. Given the uncertainties in the models of dust opacities
    and variation of the gas-to-dust ratio, $\beta$ is within the errors 
    comparable
    to the upper limit we obtain in Sect.~\ref{g11:dentemp}.

    Comparing the flux from the Bolocam Galactic Plane Survey 
    \citep[BGPS\footnote{http://irsa.ipac.caltech.edu/data/BOLOCAM\_GPS/};][]{ag11}
    1.1 mm with the PdBI 1.2 mm continuum flux we find that the interferometric
    observations filter out about 75\% of the emission.  

    To make a rough mass estimate of the core from the 3 mm data, all 
    parameters remain 
    the same as with the 1 mm data except for the flux density, $\sim$20 mJy,
    and $\kappa_{\nu} =$ 0.2~cm$^2$~g$^{-1}$ (extrapolating from Table 1 
    Column 9 in \citealp*{os94}). We obtain a core mass of 
    37~$M_{\odot}$.

   \subsubsection{Density, temperature, and spectral indeces}
\label{g11:dentemp}  
\begin{figure}
   \centering
   \includegraphics[width=8.5cm]{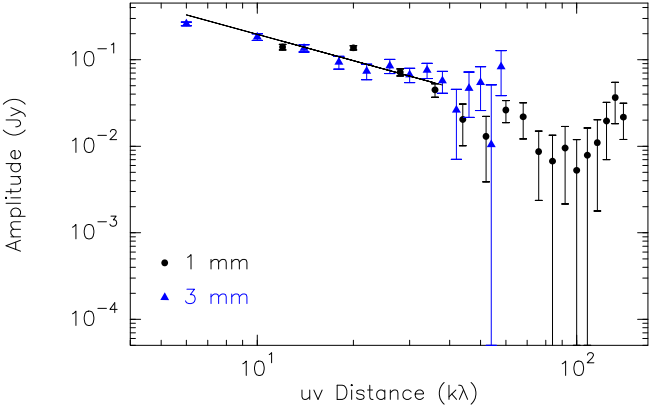}
      \caption{Continuum emission from G11.11P1 at 3.1 mm 
        (filled triangles) and 1.2 mm  (filled circles) in the Fourier domain.
        Amplitudes are averaged in bins of deprojected $u$-$v$ 
        distance from the continuum intensity peaks.  
        The 3.1 mm $u$-$v$ amplitudes have been scaled by a factor of 15.
        The straight solid line represents the least-squares fit to the data. 
        Error bars are for 1$\sigma$ statistical errors based on the 
        standard deviation of the mean of the data points in the bin.
}
         \label{Fklamall15}
   \end{figure}

    For deeper insight into the source structure, we have analyzed the 
    1 mm and 3 mm continuum data in the $u$-$v$ domain avoiding the cleaning 
    and $u$-$v$ sampling effects. Figure \ref{Fklamall15} shows the averaged 
    amplitudes averaged over a concentric annulus (from the continuum peak positions) 
    versus deprojected $u$-$v$ distance in units of wavelength. The amplitudes 
    have been averaged vectorially and the 3 mm $u$-$v$ amplitudes in Jy 
    were scaled by a 
    factor of 15 to match the 1 mm visibilities. Error bars are 
    1$\sigma$ statistical errors based on the standard deviation of the mean 
    of the data points in the bin.

    Since the index of the power law in $u$-$v$ distance is related to 
    the index of the power law in radial distance, $r$, following
    \citet{lo03}, we can write
    $F(r) \propto r^{-(p+q)+1} \rightarrow V(s) \propto s^{(p+q)-3}$,
    where $F(r)$ is the flux density in the image domain, $V(s)$ is the 
    Fourier transform of the flux density, $s$ is 
    $u$-$v$ distance and, $p$ and $q$ are the power-law indices of the 
    density ($n \propto r^{-p}$) and temperature structure 
    ($T \propto r^{-q}$), respectively.

    The least-squares fit (see Fig.~\ref{Fklamall15}) to both the 1 mm 
    and 3 mm data yields a slope of $-$1.01 $\pm$ 0.02, corresponding 
    to $p + q \sim$ 2.0. Points starting at 40 k$\lambda$, which corresponds to
    an angular scale of $\sim$6\arcsec~(0.1 pc), were excluded from 
    the fit because the signal-to-noise ratio gets very low.
    If the index $q =$  0.4  
    \citep{go74}, the density profile index is 1.6.
    
    We estimate a mean volume gas density of the core 
    ($n = \frac{M_g}{\frac{4}{3}\pi R^3\mu_{H_2} m_H}$, where $R$ is the core radius) of 
    $7 \times 10^5$~\cmmth~using the mass derived from the dust 
    observations at 1 mm (13 \msun) and 
    the geometric mean of the deconvolved linear size as the core radius
    ($R = \sqrt{(FWHM_{\rm maj}/2)\times (FWHM_{\rm nin}/2)}$ = 0.04 pc).

    The scaling factor of 15 corresponds to a spectral index 
    $\alpha =$ 3.0 $\pm$ 0.4 and we thus obtain an 
    opacity spectral index $\beta =$ 1.0 $\pm$ 0.4. 
     The derived dust opacity index corresponds, taking the upper limit, to a 
   ``normal'' interstellar dust material value. Nonetheless, if taking the 
   lower limit, the dust opacity index goes down to a value that is usually 
   found toward disks \citep[e.g.,][]{be91,na04}, which has been ascribed to 
   grain growth. Alternatively, the presence of winds/outflows (see Sect.~\ref{g11:diout}) 
   could mimic a low value of $\beta$ \citep{beu07}. Here we adopt a 
   $\beta = 1.6$ in the rest of the paper.

\subsection{PdBI: Molecular emission}\label{g11:molecular}
\begin{figure*}
   \centering
   \includegraphics[angle=-90,width=17.5cm]{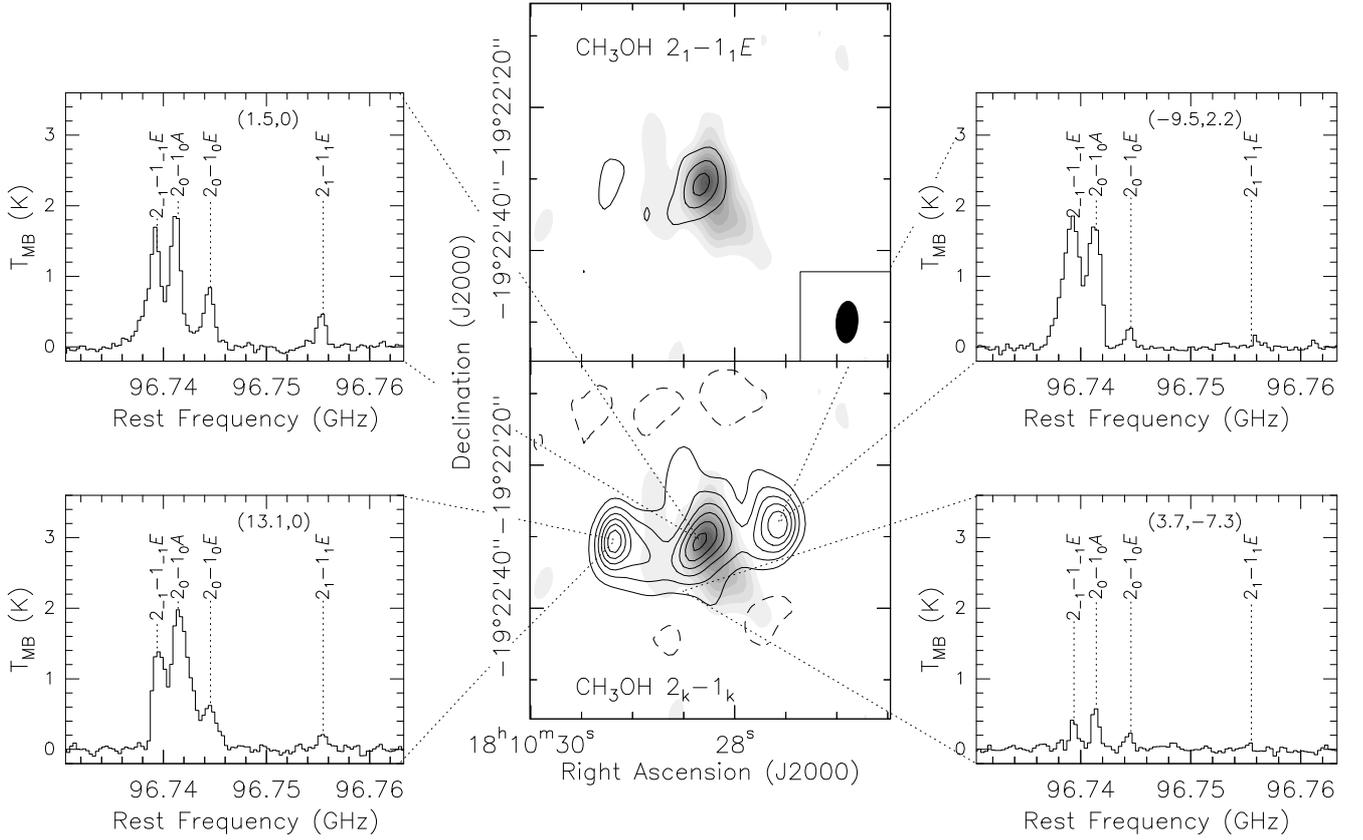}
      \caption{
        In grey-scale the 3.1 mm continuum emission of G11.11P1.
        {\it Top middle}: Contour image of the CH$_{3}$OH emission 
        integrated under the line 2$_{1} \rightarrow$ 1$_{1}E$. 
        First contour and contour spacing are 
        0.66 Jy~beam$^{-1}$~\kms~(3$\sigma$). The 
        synthesized beam (6$\farcs$0 $\times$ 3$\farcs$2; PA = 
        174\degr) is shown in the lower right corner.
        {\it Bottom middle}: Contour image of the CH$_{3}$OH emission 
        integrated under the 2$_{-1} \rightarrow$ 1$_{-1}E$, 
        2$_{0} \rightarrow$ 1$_{0}A$, and 2$_{0} \rightarrow$ 1$_{0}E$
        transition lines.
        First contour and contour spacing are 
        0.45~Jy~beam$^{-1}$~\kms~(3$\sigma$). The dashed contours show the
        negative emission (-3$\sigma$).
        Shown are also spectra at four different positions; positions are given 
        in each inset in parenthesis in units of arcsec.
              }
         \label{F3mm}
   \end{figure*}

    Spectroscopic parameters of the detected C$^{34}$S 2 $\rightarrow$ 1 and 
    CH$_3$OH 2$_{k} \rightarrow$ 1$_{k}$ and 5$_{k} \rightarrow$ 4$_{k}$ 
    transitions toward G11.11P1 are listed in Table~\ref{Teup}. 
    The interferometric integrated intensity maps, overlaid on
    their corresponding continuum emission, along with sample spectra
    are presented in Figs.~\ref{Fc34s}--\ref{F1mm}.

\begin{table}
\begin{minipage}{\columnwidth}
\caption{Molecular parameters of detected line transitions.}
\label{Teup}  
\centering
\renewcommand{\footnoterule}{}
\begin{tabular}{l l r r r} 
\hline\hline
\noalign{\smallskip}
Molecule\footnote{Rest frequencies, upper level energies and $\mu^2S$ from the Cologne Database for Molecular Spectroscopy \citep[CDMS;][]{mu01,mu05} as of December 2010.} & Transition & Frequency & $E_u$/$k$ & $\mu^2S$ \\
 & & (MHz) & (K) & (D$^2$)\\
\noalign{\smallskip}
\hline
\noalign{\smallskip}
C$^{34}$S 
& 2$\rightarrow$1                & 96412.9495   & 6.94 &7.6678\\

CH$_3$OH
&2$_{-1} \rightarrow$ 1$_{-1}~E$  &  96739.3620  &  12.55 &  1.2134 \\
&2$_{0} \rightarrow$ 1$_{0}~A$    &   96741.3750 &   6.97 &   1.6170\\
&2$_{0} \rightarrow$ 1$_{0}~E$    &   96744.5500 &  20.10 &   1.6166\\
&2$_{1} \rightarrow$ 1$_{1}~E$    &   96755.5110 &  28.03 &   1.2442\\  
&5$_{0} \rightarrow$ 4$_{0}~E$    &  241700.2190 & 47.96  &  4.0402\\
&5$_{-1} \rightarrow$ 4$_{-1}~E$  &  241767.2240 & 40.41  &  3.8824\\
&5$_{0} \rightarrow$ 4$_{0}~A$    &  241791.4310 & 34.83  & 4.0430 \\
&5$_{\pm 3} \rightarrow$ 4$_{\pm 3}~A$&241832.9100&  84.68 &  2.5775 \\
&5$_{1} \rightarrow$ 4$_{1}~E$    &  241879.0730   & 55.91  & 3.9802 \\
&5$_{-2} \rightarrow$ 4$_{-2}~E$  &  241904.1520\footnote{Blend of lines.}& 60.76  & 3.3986 \\
&5$_{2} \rightarrow$ 4$_{2}~E$    &  241904.6450$^b$   & 57.11  & 3.3558\\
\noalign{\smallskip}
\hline  
                 
\end{tabular}
\end{minipage}
\end{table}

\subsubsection{C$^{34}$S emission}\label{g11:c34s}

\begin{figure}
   \centering
   \includegraphics[width=6.7cm]{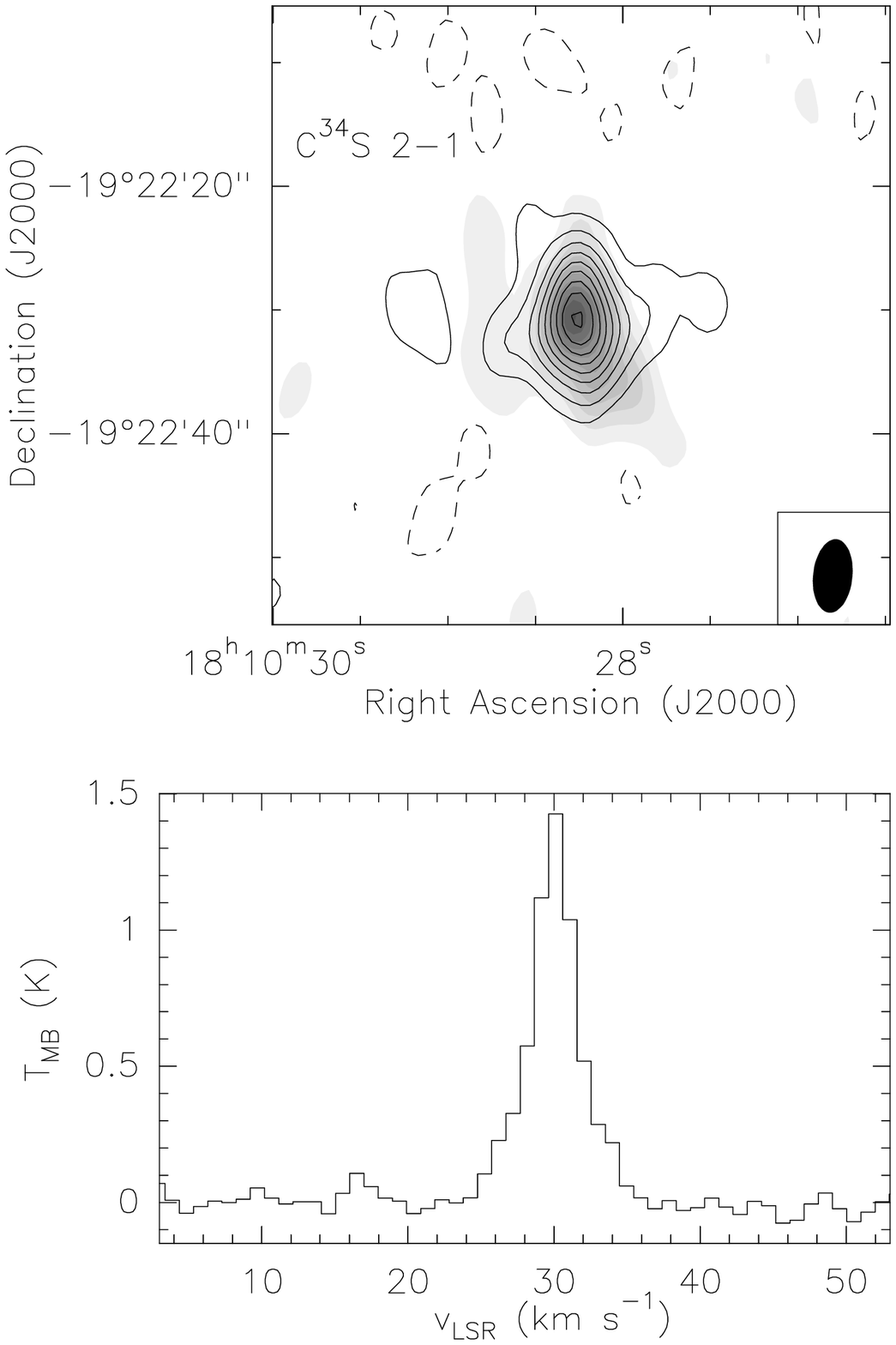} 
      \caption{
        {\it Top}:
        Contour image of the emission integrated under the 
        C$^{34}$S 2$ \rightarrow$ 1 line, in the velocity range from
        22.4 to 37.9 \kms~, overlaid on the
        grey-scale 3.1 mm continuum emission of G11.11P1.
        First contour and contour spacing are
        0.09~Jy~beam$^{-1}$~\kms~(3$\sigma$), the dashed contours show
        the negative emission (-3$\sigma$). The 
        synthesized beam (6$\farcs$0 $\times$ 3$\farcs$2; PA = 
        176\degr) is shown in the lower right corner.
        {\it Bottom}:  Spectrum of C$^{34}$S 2$ \rightarrow$ 1
        toward the peak position on the integrated intensity map: 
        R.A. = 18\fho10\fmi28\fs24, Decl. = 
        -19\degr22\arcmin30\farcs5 (J2000). 
      }
         \label{Fc34s}
   \end{figure}

    The integrated intensity map of the C$^{34}$S 2 $\rightarrow$ 1
    line, is shown in the {\it top} panel of Fig.~\ref{Fc34s}. 
    This emission is spatially coincident with the 3 mm
    continuum; moreover, some weaker emission (at 3$\sigma$ level) extends along the E-W direction. 
    Observed parameters toward the peak position in the integrated 
    intensity map are presented in Table~\ref{Ttrans}.

\begin{table*}
\begin{minipage}{\textwidth}
\caption{Observed molecular line parameters from Gaussian fits.}
\label{Ttrans}
\centering
\renewcommand{\footnoterule}{}
\begin{tabular}{l l  c c c c} 
\hline\hline
\noalign{\smallskip}
Molecule
& Transition
           & v$_{\rm LSR}$  & Peak $T_{\rm {MB}}$     & $\Delta$V (FWHM)  & $\int T_{\rm {MB}} d$V \\
& &  (\kms) & (K)           &  (\kms)& (K~\kms) \\ 
\noalign{\smallskip}
\hline
\noalign{\smallskip}
\multicolumn{6}{c}{Offset (0\arcsec,\,0\arcsec)}\\
C$^{34}$S
& 2$ \rightarrow$ 1              &  29.90 (0.04) &  1.31 (0.04)  & 4.1 (0.1)& 5.6 (0.1)\\
          
\noalign{\smallskip}
\hline 
\noalign{\smallskip}                 
 \multicolumn{6}{c}{Offset (1\farcs5,\,0\arcsec)}\\                                
CH$_3$OH                         
& 2$_{-1} \rightarrow$ 1$_{-1}~E$ &  29.85 (0.05) & 1.49 (0.04)   &  5.1 (0.2)  & 8.0 (0.2) \\
& 2$_{0} \rightarrow$ 1$_{0}~A$   &  29.72 (0.04) & 1.89 (0.04)   &  3.6 (0.1)  & 7.3 (0.2) \\
& 2$_{0} \rightarrow$ 1$_{0}~E$   &  29.87 (0.09) & 0.78 (0.04)   &  4.4 (0.2)  & 3.6 (0.1) \\
& 2$_{1} \rightarrow$ 1$_{1}~E$   &  29.79 (0.01) & 0.49 (0.04)   &  3.1 (0.2)  & 1.6 (0.1) \\
                                     
& 5$_{0} \rightarrow$ 4$_{0}~E$   &  29.71 (0.20)  & 0.70 (0.04)   &   4.1 (0.5) & 3.1 (0.3) \\
& 5$_{-1} \rightarrow$ 4$_{-1}~E$ &  29.92 (0.09)  & 1.48 (0.04)   &   4.1 (0.2) & 6.5 (0.3) \\
& 5$_{0} \rightarrow$ 4$_{0}~A$   &  30.00 (0.05)  & 1.80 (0.04)   &   4.3 (0.2) & 8.3 (0.3) \\
& 5$_{1} \rightarrow$ 4$_{1}~E$   &  29.78 (0.41)  & 0.33 (0.04)   &   3.9 (1.3) & 1.4 (0.3) \\
\noalign{\smallskip}
\hline 
\end{tabular}
\tablefoot{CH$_3$OH 2$_{k} \rightarrow$ 1$_{k}$ and 5$_{k} \rightarrow$ 4$_{k}$ line parameters listed in this table were used in the rotational diagram method described in the text. The CH$_3$OH 5$_{k} \rightarrow$ 4$_{k}$ transitions have been convolved to the 2$_{k} \rightarrow$ 1$_{k}$ resolution.}
\end{minipage}
\end{table*}

    {\bf Virial mass.} The virial mass of a core with a 
    power-law density distribution is 
    $M_{\rm vir} = k_1 \sigma_{\rm v}^2R/G$ \citep{mac88} where 
    $k_1 = (5-2p)/(3-p)$, $\sigma_{\rm v}$ is the three-dimensional 
    root-mean-square velocity which is related to the FWHM line width, 
    $\Delta{\rm V}$, through $\sigma_{\rm v}^2 = (3/8{\rm ln2})\Delta{\rm V^2}$,
    $R$ is the core radius and $G$ is the Gravitational constant. For $p = 1.6$
    (see Sect.~\ref{g11:dentemp}), the virial mass can be expressed as
\begin{equation}
      M_{\rm vir} \simeq~161 \times 
\left(\frac{R}{\rm pc}\right)~\left(\frac{\Delta{\rm V}}{\rm km~s^{-1}}\right)^2 \label{emvir}\,~\msun.
   \end{equation}
    A virial mass of $\sim$135~\msun~is calculated when using 
    $\Delta{\rm V} = 4.1~\kms$ (see Table~\ref{Ttrans})
    of the C$^{34}$S optically thin line and $R =$ 0.05 pc, which 
    corresponds the geometric mean of the deconvolved linear size.

    {\bf Virial parameter.} The virial parameter \citep{ber92} for a spherical 
    cloud is defined as $\alpha_{\rm vir} \equiv M_{\rm vir}/M_{\rm tot}$, where 
    $M_{\rm tot}$ is the total mass. If we assume that 
    $M_{\rm tot}= M_{\rm g} = 37~\msun$, then $\alpha_{\rm vir} = 3.6$. 
    
    It is worth noting that the C$^{34}$S line width could be affected 
      by unbound motions, e.g., 
      outflows, partially broadening the line and thereby increasing the virial
    mass estimate significantly and the virial parameter as well.

    {\bf Jeans mass.} Following \citet{sta05}, we can calculate the Jeans mass 
\begin{equation}
      M_J = \frac{ma^3}{\rho^{1/2}~G^{3/2}}
\simeq~1.0 \times 
\left(\frac{T}{10~{\rm K}}\right)^{3/2}\left(\frac{n}{10^{4}~{\rm cm^{-3}}}\right)^{-1/2} \label{emj}\,~\msun,
   \end{equation}
   where $a$ is the isothermal sound speed and $\rho$ is the mass density.
   For $n = 7 \times 10^5$~\cmmth~(see Sect.~\ref{g11:dentemp}) and $T =$ 60~K,
   we obtain a Jeans mass of $\sim$1.8~\msun.
   We find a gas mass ($M_g$) to the Jeans mass ($M_J$) ratio 
     larger than unity;  
   other interferometric studies 
   \citep{rath08,zh09} toward IRDCs report $M_g/M_J > 1$
   as well.

   From our interferometric observations at 1mm, and taking a flux density 
   equal to the 3$\sigma$ detection level, we derive a mass sensitivity limit
   of 0.4~\msun. Comparing this value with the Jeans mass, we note that our 
   mass sensitivity limit is good enough to detect fragments of the order of the 
   Jeans mass.


\subsubsection{CH$_{3}$OH emission}\label{g11:ch3oh}
    Figure~\ref{F3mm} shows the integrated intensity map of the
    CH$_{3}$OH $2_{k} \rightarrow 1_{k}$ quartet of lines
    ({\it bottom middle}). This emission presents 
    three maxima; one of them is spatially coincident with the 1 and 3 mm 
    continuum emission which peaks at (1\farcs5,\,0\arcsec), the other 
    two are located at (13\farcs1,\,0\arcsec) and ($-$9\farcs5,\,2\farcs2).
    We plot sample spectra at four different positions, including the three
    maxima and one position on the envelope (3\farcs7,\,$-$7\farcs3), in
    order to compare line profiles, absolute line strengths, and the relative
    strengths among lines.
    The high excitation CH$_3$OH $2_{1} \rightarrow 1_{1}E$ transition 
    that arises from a smaller region is also shown ({\it top middle}).

    Line widths at and close to the three peak positions are, in average 
    $\sim$5~\kms~with line profiles showing red- and/or blueshifted 
    ``wings'', while in the outer parts of the envelope, e.g., at the 
    (3\farcs7,\,$-$7\farcs3) position, the 
    line width is narrower ($\sim$2 \kms; see spectra in Fig.~\ref{F3mm}).

\begin{figure}
   \centering
   \includegraphics[width=6.8cm]{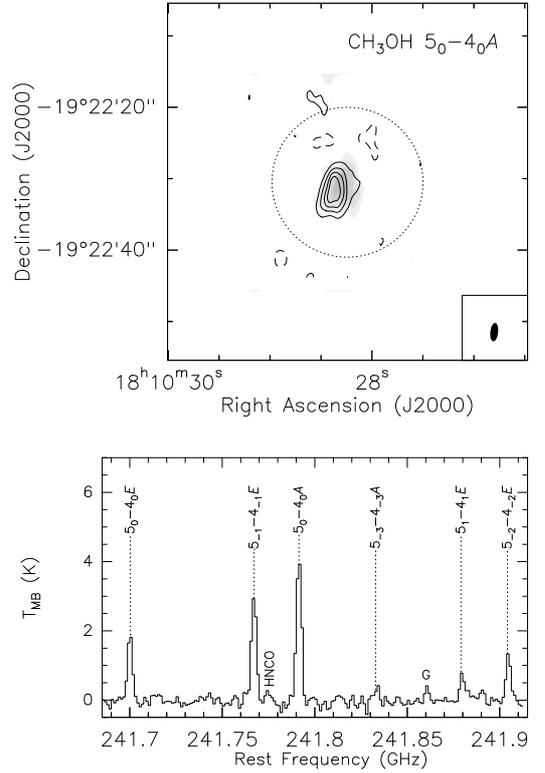}
      \caption{
        {\it Top}:
        Contour image of the emission integrated 
        under the 5$_{0} \rightarrow$ 4$_{0}A$ 
        transition line, overlaid on the grey-scale 1.2 mm continuum 
        emission of G11.11P1. First contour and contour spacing are 0.16
        Jy~beam$^{-1}$~\kms~(3$\sigma$); the dashed contours show the
        negative emission (-3$\sigma$). The 
        synthesized beam (2\farcs6 $\times$ 1\farcs1; PA = 
        174\degr) is shown in the lower right corner.
        The dotted circle indicates the interferometer primary beam (21\arcsec)
        at this frequency.
        {\it Bottom}: Spectrum of CH$_3$OH 5$_{k} \rightarrow$ 4$_{k}$
        towards the position: 
        R.A. = 18\fho10\fmi28\fs40, Decl. = 
        -19\degr22\arcmin31\farcs9 (J2000). 
         The HNCO 11$_{0,11} \rightarrow$ 10$_{0,10}$ line 
        (at 241774.0320 MHz with $E_u/k$ = 69.67 K and 
        $\mu^2S$ = 27.4590 D$^2$) is shown 
        as well. The ``G'' shows an artifact produced in the central
        channels of a correlator unit due to Gibbs effect.
               }
         \label{F1mm}
   \end{figure}

    In Fig.~\ref{F1mm} we present the integrated intensity map of the
    strongest line CH$_{3}$OH 5$_{0} \rightarrow$ 4$_{0}A$. The
    emission distribution is as extended as that of the 1 mm continuum
    and the CH$_3$OH 2$_{1} \rightarrow$ 1$_{1}E$ emissions.

   {\bf Local thermodynamic equilibrium (LTE) analysis.} 
   We derived the rotational temperatures and methanol column densities by 
   using the rotational diagram method \citep[e.g.,][]{cu86,go99}. 
   Briefly, for an optically thin line, the column density 
   in the upper level ($N_u$) can be written as
   $N_u = 3kW/(8\pi^3 \nu S \mu^2)$,
   where $k$ is the Boltzmann constant, $W$ is the integrated intensity of a 
   line ($\int T_{\rm {MB}} d$V), $\nu$ the frequency, $S$ the line strength, 
   $\mu$ the dipole moment. Under LTE conditions, the population
   of each level follows a Boltzmann law at the gas temperature $T$ by
   $N_u = (N/Q)g_u e^{-E_u/kT}$,
   where $N$ is the total column density, $Q$ the partition function, and
   $g_u$ and $E_u$ are the statistical weight and the energy of the upper
   level, respectively. We then obtain

\begin{equation}
{\rm {log}}\left(\frac{N_u}{g_u}\right) = {\rm {log}}\left(\frac{N}{Q}\right) - \frac{{\rm {log}}e}{T}\frac{E_u}{k} \label{erot}\,.
\end{equation}

\begin{figure}
   \centering
   \includegraphics[angle=-90,width=8.5cm]{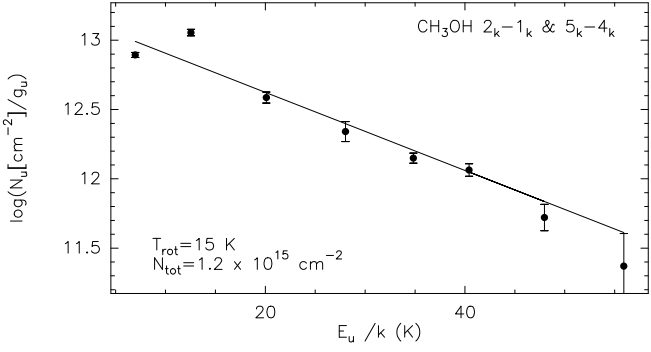}
      \caption{Rotational diagram of CH$_3$OH (2--1) and (5--4) transition lines 
towards the (1\farcs5,\,0\arcsec) position.
The straight line is a
        least-squares fit to the data corresponding to the rotational
       temperature and column densities indicated in the bottom left.
       Error bars are for 1$\sigma$ uncertainties.
             }
         \label{Frot}
   \end{figure}

   Using Eq.~\ref{erot}, we plot $\rm{log} (N_u/g_u)$ versus $E_u/k$ 
   (see Fig.~\ref{Frot}) and make a least-squares fit for $T$ and $N$ 
   taking into account both CH$_3$OH 2$_{k} \rightarrow$ 1$_{k}$ and 
   $5_{k} \rightarrow 4_{k}$ bands. The partition function of CH$_3$OH
   is $Q = 2[\frac{\pi~(kT)^3}{h^3ABC}]^{1/2}$ \citep{tur91}, where
   $h$ is the Planck constant and $A$, $B$, and $C$ are the rotation constants
   \citep[CDMS;][]{mu01,mu05}. A $Q = 1.23~T^{1.5}$
   was used in the calculations. The CH$_3$OH 
   $5_{k} \rightarrow 4_{k}$ data were smoothed to the resolution of 
   the $2_{k} \rightarrow 1_{k}$ data. CH$_3$OH transition lines that were used
   in this fit are listed in Table \ref{Ttrans}.

   To estimate the CH$_3$OH fractional abundance (relative to H$_2$), 
   $X_{\rm CH_3OH}$, we calculate
   the beam-averaged H$_2$ column density, $N_{\rm H_2}$, following the expression 
    \begin{equation}
      N_{\rm H_2} = \frac{I_{\nu}^{\rm peak}~R_{\rm gd}}{\kappa_{\nu}~B_{\nu}(T_{\rm d})~\Omega~\mu_{\rm H_2}~m_{\rm H}} \label{enh2}\,~\cmmtw,
   \end{equation} 
   where $I_{\nu}^{\rm peak}$ is the
   peak intensity in Jy/beam, $\mu_{\rm H_2} =$ 2.8 \citep{ka08} is the molecular 
   weight per hydrogen molecule, and 
   $\Omega = (\pi \theta_{\rm min} \theta_{\rm maj})/(4 \rm{ln2}) $ is the 
   beam solid angle of an elliptical Gaussian beam with minor and major axes 
   $\theta_{\rm min}$ and $\theta_{\rm maj}$, respectively. 

   We obtained a rotational 
   temperature of 15~K and a CH$_3$OH column density, $N_{\rm CH_3OH}$, 
   of $1.2 \times 10^{15}$~\cmmtw~towards the (1\farcs5,\,0\arcsec) 
   peak position.
   Since the lines are likely sub-thermally excited, we take a dust temperature
   $T_{\rm d} =$ 60 K (see LVG analysis) and  obtain a 
   $N_{\rm H_2} = 4.6~\times 10^{22}$~\cmmtw~(from our 1 mm continuum data convolved to the 3 mm data) and a $X_{\rm CH_3OH}$  
   of 2.6~$\times$~10$^{-8}$.

   At the (13\farcs1,\,0\arcsec) position, we get a $X_{\rm CH_3OH}$ = 
   3.7~$\times$~10$^{-9}$, when using 
   $N_{\rm CH_3OH}$ = 2.9~$\times$~10$^{14}$ \cmmtw~and 
   $N_{\rm{H_2}}$ = 7.7~$\times$~10$^{22}$  \cmmtw. A rotational temperature of
   13~K was calculated. At the ($-$9\farcs5,\,2\farcs2) position, the
   values for the $X_{\rm CH_3OH}$, $N_{\rm CH_3OH}$, and 
   $N_{\rm{H_2}}$ are: 5.6~$\times$~10$^{-9}$, 3.8~$\times$~10$^{14}$ \cmmtw, and
   6.8~$\times$~10$^{22}$ \cmmtw, respectively. For the 
   ($-$9\farcs5,\,2\farcs2) peak, we fixed $T$ = 15 K because of low 
   signal-to-noise of one of the three lines used in the fit. 
   $N_{\rm CH_3OH}$ for both peaks (13\farcs1,\,0\arcsec)~and ($-$9\farcs5,\,2\farcs2) were calculated from the 3 mm 
   interferometric map only and $N_{\rm H_2}$ from the 
   SCUBA 850 $\mu$m dust continuum; the interferometric observations were 
   convolved to the SCUBA beam size (14\arcsec).

\begin{figure}
   \centering
   \includegraphics[width=6.5cm]{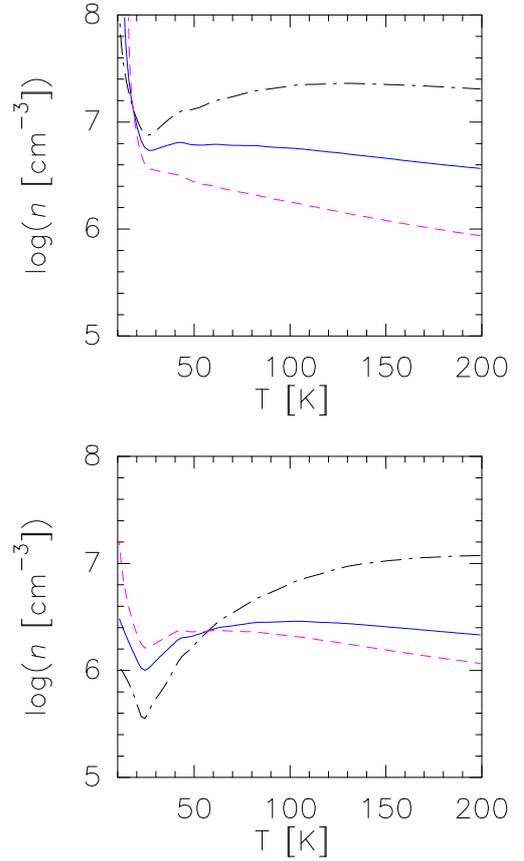}
      \caption{Results of statistical equilibrium calculations for 
        CH$_3$OH-$E$. The 5$_0 \rightarrow 4_0/5_{-1} \rightarrow 4_{-1}$ 
        (black dash-dotted line),
        5$_{1} \rightarrow 4_{1}/5_{-1} \rightarrow 4_{-1}$ 
        (blue solid line), and 5$_1 \rightarrow 4_1/5_0 \rightarrow 4_0$ 
        (pink dashed line) observed integrated intensity line ratios 
        towards the (1\farcs5,0\arcsec)~position in logarithmic 
        scale are shown as function of H$_2$ density and temperature at 
        $N_{\rm CH_3OH}/\Delta$V $= 10^{14}$~\cmmtw/(\kms)~({\it top}) 
        and $N_{\rm CH_3OH}/\Delta$V $= 10^{15}$ \cmmtw/(\kms)~({\it bottom}).
                     }
         \label{Fcolden}
   \end{figure}

   {\bf Large velocity gradient (LVG) analysis.} 
   \citet{le04} found several ratios, from some CH$_3$OH-$E$
   transitions, to be calibration-independent tracers of density in the
   1 and 3 mm bands. We have used their LVG 
   calculations and have plotted the following 
   integrated intensity line ratios:
   $5_{0} \rightarrow 4_{0}/5_{-1} \rightarrow 4_{-1}$, 
   $5_{1} \rightarrow 4_{1}/5_{-1} \rightarrow 4_{-1}$, and
   $5_{1} \rightarrow 4_{1}/5_{0} \rightarrow 4_{0}$,
   in a temperature-density ($T$-$n$) plane. In Fig.~\ref{Fcolden}, the 
   observed ratios, towards the (1\farcs5,0\arcsec)~position, are represented by the black dash-dotted line, blue solid 
   line, and pink dashed line.
   The $2_{k} \rightarrow 1_{k}$ quartet of lines in the 3 mm band were not 
   used in the analysis because they are blended.
   
   Given the results for CH$_3$OH column densities in the LTE analysis, the explored LVG models for the excitation of methanol run for temperatures 
   in the range of 10--200 K, for densities 10$^5$--10$^8$~\cmmth, 
   and for two CH$_3$OH column densities per line width 
   10$^{14}$ and 10$^{15}$~\cmmtw/(\kms).    

  For an average line width of 4~\kms,
   this corresponds to $N_{\rm CH_3OH-E} = 4 \times 10^{14}$~\cmmtw~and  4 $\times 10^{15}$~\cmmtw, respectively. 
   Depending on the column density, two possible solutions are found: 
   at $N_{\rm CH_3OH-E} = 4 \times 10^{14}$~\cmmtw, the three integrated
   intensity line ratios intercept at $T =$ 17 K and $n$ = 10$^7$~\cmmth. On the other hand, at 
   $N_{\rm CH_3OH-E} = 4 \times 10^{15}$~\cmmtw, the interception is at 
   $T$ = 60 K and $n = 2 \times 10^6$~\cmmth~(see Fig.~\ref{Fcolden}).
   The methanol column density inferred in this case is consistent with the LTE 
   analysis within a factor of 3,
   but the intensities in this model are one order of magnitude higher than the observed line intensities.
   In other words, to fit the observations with the $N_{\rm CH_3OH-E} = 4 \times 10^{15}$~\cmmtw~model, 
   we need a beam filling factor of 0.1 resulting in a more
   compact source (with a size of 0\farcs5 corresponding to 0.009 pc or 1800 AU), even smaller than the one
   modelled by \citet{pi06a} and \citet{le07b}.

\begin{figure*}
   \centering
   \includegraphics[angle=-90,width=17.7cm]{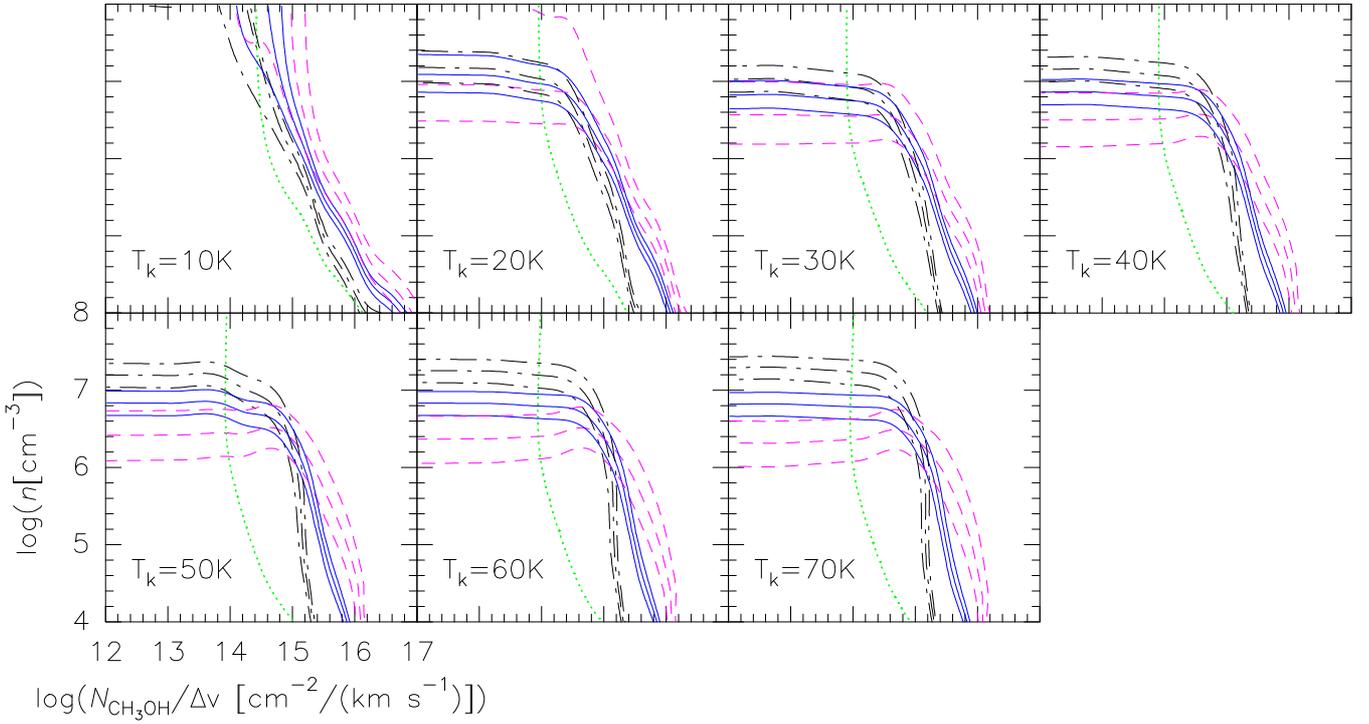}
      \caption{Results of statistical equilibrium calculations for 
        CH$_3$OH-$E$. Observed integrated intensity line ratios are color-coded
        as in Fig.~\ref{Fcolden} and plotted in logarithmic scale 
        (along with their 1$\sigma$ values)
        as function of H$_2$ density and $N_{\rm CH_3OH}/\Delta{\rm V}$ at different
        temperatures (10--70~K). The green dotted line represents the peak
        main brightness temperature of the 5$_{0} \rightarrow 4_{0}E$ 
        transition.}
         \label{FnN}
   \end{figure*}

   To gain more insight into the possible $T$-$n$ degeneracy,
   in Figure~\ref{FnN}, we show $n$ vs. $N_{\rm CH_3OH}$ plots, at different
   temperatures (10-70 K), where the color-coding for the ratios 
   (along with their 1$\sigma$ values) are the same 
   as in Fig.~\ref{Fcolden}; additionally, we show in each panel the peak main 
   beam brightness temperature of the 5$_{0} \rightarrow 4_{0}E$ transition 
   (green dotted line). 
   The parameters obtained in the regions in the left side of the green 
   dotted line 
   imply beam filling factors greater than 1, which is not possible. 
   In each panel, the areas of interest are those that lie in the right
   side of the green line and, at the same time, are delimited by the 
   integrated intensity ratios (taking into account their 1$\sigma$ values). 
   We see that, for $T$ in the range 30-70 K, the $n$ remains constant up to 
   $N_{\rm CH_3OH}$~$\sim$~4~$\times$ 10$^{14}$~\cmmtw, while for the very
   low $T$ (= 10 K), there is no solution.

   Given our density estimate from the dust continuum, we view the low $T$,
   high $n$ solutions with a large filling factor as unlikely.
   Therefore, although we cannot strongly constrain the physical parameters of 
   the gas emitting methanol from the LVG analysis because of the limited 
   number of observed transitions, we can exclude the low temperature solution 
   and put a lower limit to the 
   kinetic temperature of the gas to 30~K. This result is in agreement with  
  the detection of methanol maser from this region and with a previous 
  measurement of $T=60$~K from NH$_3$ observations. Therefore, we adopt 
  $T=60$~K as kinetic temperature at the peak of the continuum emission.

\subsection{PdBI+30m: extended CH$_3$OH  emission}\label{smerged}

\begin{figure}
   \centering
 \includegraphics[angle=-90,width=8cm]{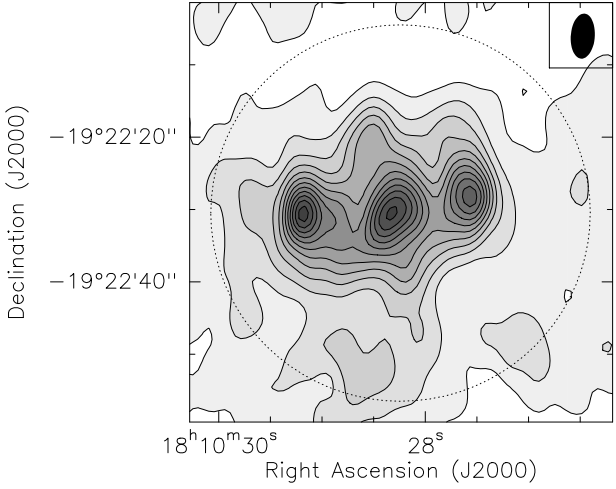}
     \caption{
        CH$_3$OH 2$_{k} \rightarrow$ 1$_{k}$ $v_{t} = 0$ integrated intensity 
        map combining PdBI and 30m data. The emission is integrated over 
        the 2$_{-1} \rightarrow$ 1$_{-1}E$, 2$_{0} \rightarrow$ 1$_{0}A$,
        and 2$_{0} \rightarrow$ 1$_{0}E$
        transitions. First contour and contour spacing are 0.39
        Jy beam$^{-1}$ km~s$^{-1}$ (3$\sigma$), the dashed contours show the
        negative emission (-3$\sigma$). The synthesized beam 
        (6$\farcs$2 $\times$ 3$\farcs$3; PA = 176\degr) is shown in the upper 
        right corner.
        The dotted circle indicates the interferometer primary beam (52\arcsec). 
      }
         \label{Fmerged}
   \end{figure}

    A comparison of the interferometric data with the single-dish observations
    shows that only $\sim$30\% of the CH$_{3}$OH 
    2$_{k} \rightarrow$ 1$_{k}$ measured
    with the 30 m telescope is imaged in the PdBI observations.
    The combined PdBI and 30 m telescope data recover almost 100 \%
    of the flux at the central (1\farcs5,0\arcsec)~position.
    In Fig.~\ref{Fmerged}, we show the resultant map after combining the 
    CH$_3$OH 2$_{k} \rightarrow$ 1$_{k}$ single-dish observations with the
    corresponding PdBI data. The methanol emission is extended already in 
    the interferometric
   observations (mapping angular scales as large as 32\farcs5) and 
   confirmed to be so in the resultant merged map with the
   single-dish data. It is important to note that the emission is extending 
   over 1 pc scale in the merged map (see Fig.~\ref{Fmerged}). 
   Similarly, silicon monoxide (SiO) was found to be widespread over
    a 2 pc scale in the IRDC~G035.39-00.33 \citep{jim10}.
    The 2$_{1} \rightarrow$ 1$_{1}E$ transition was not detected 
    in the single-dish observations, which is expected since
    this emission does not appear to be extended in our interferometric data.

    As for the 1mm data, the missing flux 
    in the interferometric observations is $\sim$70\%.
    In the 1 mm single-dish observations,
    two of the strongest transitions showed up, namely: 
    the 5$_{-1} \rightarrow$ 4$_{-1}E$ and
    the 5$_{0} \rightarrow$ 4$_{0}A$ lines. 
    Both lines are above 3$\sigma$ noise level at 6 positions out of 
    144 observed positions.

    How is the derivation of column densities and rotational temperatures
    affected by missing flux? Using the combined data at 3 mm only, the slope, 
    i.e.,
    the rotational temperature, in the rotational diagram is not 
    affected, whithin the errors, while CH$_3$OH column densities are higher 
    by a factor of two.

\section{Discussion}\label{g11:disc}

\subsection{Properties of G11.11P1 compared to other cores}\label{g11:prop}

   Despite the fact that our interferometric observations have enough
   sensitivity to resolve the Jeans mass (1.8~\msun) at 1 mm, the angular 
   resolution (2\farcs6$\times$1\farcs1) is smaller than but very close to the 
   Jeans length of 
   3\farcs2 (0.05~pc). Therefore, it is difficult to conclude whether 
   G11.11P1 does fragment or not into condensations (referred as substructures 
   within the cores) as is often found when zooming into these cores 
   \citep[e.g.,][]{zh09,bo10}. If G11.11P1 does not fragment would mean that it has
   already accumulated most of its mass that will go into 
   the final star even though $\sim$75 \% of the (dust) emission is 
   filtered out by the interferometer. With higher angular resolution observations,
   it will be possible to support either possibilities.

  The virial parameter of G11.11P1 obtained by \citet{pi06a}, from single-dish 
   NH$_3$ and dust continuum (at 870~$\mu$m) observations, is unity 
   on the scale of $\sim$0.9 pc. Although, in the present work, we find
 that $M_{\rm vir}$ is much larger than the $M_g$ on the scale of $\sim$0.1 pc
   which would indicate that
   gravity is not the dominant force 
   \citep{ber92}. We caution, however, 
   that the high value of the virial parameter ($\alpha_{\rm vir}=3.6$) could be due to the missing
   flux in the interferometric data
   and to the fact that the virial mass is estimated from the C$^{34}$S 
   line width, whose broadening is likely in part produced by outflows.
   In fact, \citet{pi06b} reported an increase in line widths, 
   from 1.29~\kms~in the NH$_3$ (1,1) line to 3.2~\kms~in the NH$_3$ (3,3) 
   data. The increase in NH$_3$ line width toward the higher $(J,K)$ transitions 
   suggest a strong influence of star forming activity on the line width at 
   smaller scales. The broad line width in the C$^{34}$S from PdBI is consistent 
   with that notion.

    
     Based on the SCUBA 870 $\mu$m column density, we estimated a
     surface density of $\Sigma = 0.4$~g \cmmtw~for G11.11P1,
  which is close 
   to the thresholds of 0.7--1.5~g 
   \cmmtw~(or 3\,320--7\,110 \msun~pc$^{-2}$) required 
to form stars of 10--200~\msun~\citep{kru08}. It is important to point 
   out that these authors were 
   interested in clouds where no massive stars have yet formed while G11.11P1
   shows already signposts of star formation activity.

   We have calculated a density profile power-law index of 1.6 from interferometric
   observations at 1 and 3 mm; similar values, in the range between 1.5 and 2, 
   for cores in the IRDC G28.34+0.06, have been reported by \citet{zh09} and 
   for embedded protostellar sources in the protocluster 
   IRAS~05358+3543 \citep{beu07} as well.

\subsection{The disk/outflow system?}\label{g11:diout}
   
   \citet{pi06b} found a velocity gradient of the CH$_3$OH 
     maser emission and explained it as the maser spots being located
   in a Keplerian disk. Interestingly, the spread of the maser components is 
   in the North-South direction which is perpendicular to 
   the CH$_3$OH $2_{k} \rightarrow 1_{k}$ emission that we propose is
   originated by an outflow (in the East-West direction).
   The spread of the masing spots is translated into a radius of $\sim$450 AU
   for a hypotetical disk in G11.11P1. 
   Note that candidate disks in high-mass protostars have masses in the 
   range 0.2--40~\msun~and radii of 500--10\,000 AU \citep{ces06}. Our 
   interferometric data set is capable of imaging the biggest disk structures,
   but not those inferred in \citet{pi06b}. 
   
   G11.11P1 has been cataloged by \citet{cy08} as a ``possible'' massive young 
   stellar 
   object (YSO) outflow candidate. This categorization was based on the 
   angular extent of the extended excess 4.5 $\mu$m emission, i.e., 
   the extent of green emission in a three-color RGB image.  

\begin{figure}
   \centering
   \includegraphics[width=7cm]{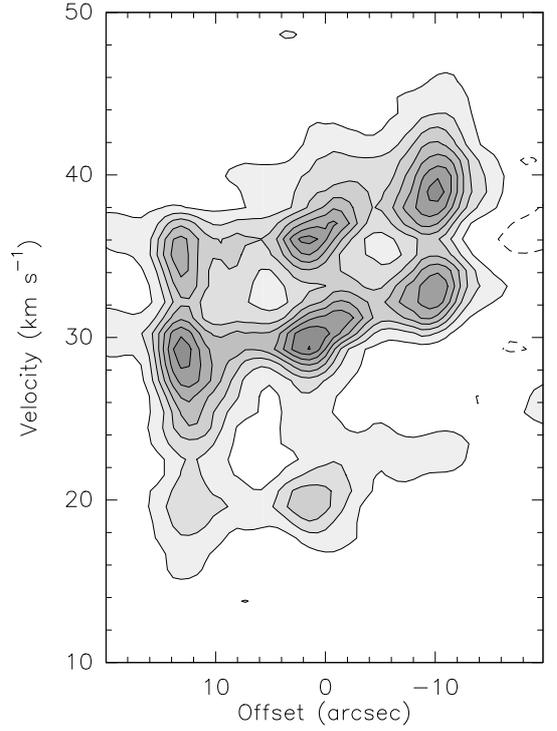}
      \caption{Position-velocity plot of the methanol 
        2$_{k} \rightarrow$ 1$_{k}$  lines, cutting along the PA=96$^\circ$
        on the map shown in the {\it middle bottom} panel of Fig.~\ref{F3mm}.
        Shown are the transitions: 2$_{-1} \rightarrow$ 1$_{-1}E$,
        2$_{0} \rightarrow$ 1$_{0}A$, and 2$_{1} \rightarrow$ 1$_{1}E$.
        The offset is measured positive towards East and from the
        pointing center of the observations. 
        First contour is 0.018 in steps of 0.036 Jy beam$^{-1}$, the
        dashed contours show the negative emission (-0.018 Jy beam$^{-1}$).
              }
         \label{Fch3ohpv}
   \end{figure}

   More evidence pointing to the presence of outflows comes from the
   non-Gaussian methanol line profiles \citep[see Fig.~\ref{F3mm} and][]{le07b}.
   In Fig.~\ref{Fch3ohpv}, we analyze the molecular emission in a 
   position-velocity plot and note that there are ``wings'' of redshifted 
   emission towards negative offsets and blueshifted emission towards positive 
   offsets (see also Fig.~\ref{F3mm}). Moreover, we find a 
   gradient of $\sim$4 \kms, in the east-west direction, for the 
   CH$_3$OH 2$_{k} \rightarrow$ 1$_{k}$ 
   emission and speculate, it is due to an outflow(s)-cloud interaction.  
   The C$^{34}$S 2$\rightarrow$1 and 
   CH$_{3}$OH 5$_{k} \rightarrow$ 4$_{k}$ lines are detected 
   mainly towards the central peak/dust emission, and no velocity gradient 
   is found.

   Another possibility is that the CH$_3$OH 2$_{k} \rightarrow$ 1$_{k}$ emission
   is actually tracing a toroid \citep[see][]{ces07} like in the 
   interferometric observations of CH$_3$OH $6_{0} \rightarrow 5_{0}$  towards 
   the IRDC~18223-3 \citep{fal09}. However, the gradient of the maser spots
   is orthogonal and not aligned with the gradient of the methanol emission
   we present in Fig.~\ref{Fch3ohpv}.

\subsection{CH$_3$OH abundances}
  \label{g11:abundances}

   Due to the fact that methanol molecules can be sub-thermally excited 
   \citep{ba95}, the temperatures derived from the rotational diagram 
      are low ($\sim$15 K) even at the central peak where masers have been 
   detected. To overcome this issue we have used the LVG method and 
   found that a higher temperature is more reliable.

   Purely gas-phase model calculations predict $X_{\rm CH_3OH}$ to be of 
   the order 10$^{-13}$--10$^{-10}$ \citep{gar06}, which cannot reproduce 
   the high values we report in this paper. \citet{gar06}
   conclude that the production of methanol is carried out on the surfaces 
   of dust grains followed by desorption into the gas.
   Then the $X_{\rm CH_3OH}$ values in G11.P11    
   could be produced by desorption of icy mantles on the dust grains.

   As found in Sect.~\ref{g11:ch3oh}, the highest $X_{\rm CH_3OH}$ in G11.11P1 
   is at the central position ($X_{\rm CH_3OH} \sim 3~\times~10^{-8}$), and it 
   decreases by more than one order of magnitude at the other two peaks, where
   the $X_{\rm CH_3OH}$ $(\sim 4-6 \times 10^{-9})$; 
   the same trend was found also by LVG-modeling of
   single-dish data \citep{le07b}. Our abundances are
   enhanced compared to dark clouds values 
   \citep[$X_{\rm CH_3OH} \sim 10^{-10}-10^{-9}$;][]{fri88}, also supporting
   the idea that methanol in G11.11P1 is forming mainly through non-thermal 
   desorption by the presence of outflow(s).

\section{Summary}
We have performed 
        continuum and line observations with the PdBI towards the IRDC
        core G11.11P1. Our main results can be summarized as follows:
   \begin{itemize}
      \item The analysis of the mm continuum maps provides a very detailed 
        physical structure of the core. 

      \item Evidence of extended 4.5 $\mu$m emission, ``wings'' in
        the CH$_3$OH~$2_{k} \rightarrow 1_{k}$ spectra, and CH$_3$OH abundance enhancement point
      to the presence of an outflow in the East-West direction.

       \item We find a 
   gradient of $\sim$4 \kms~for the CH$_3$OH~2$_{k}~\rightarrow$~1$_{k}$ 
   emission, which we interpret as being produced by an outflow(s)-cloud interaction.

      \item The fitting results show
    enhanced methanol fractional abundance ($\sim$3~$\times$~10$^{-8}$) at 
  the central peak with respect to the other two maxima where the methanol 
  abundance is lower by about an order of magnitude 
  ($\sim$4-6 $\times$ 10$^{-9}$).

   \end{itemize}

\begin{acknowledgements}
We acknowledge the IRAM staff at Pico Veleta
and Plateau de Bure for carrying out the observations.
L.G. thanks 
A. Castro-Carrizo and J. M. Winters for their help during the 
PdBI data reduction. We thank B. Parise for her evaluation on the manuscript
and the referee for providing helpful comments and suggestions.
    L.~G. was supported for this research through a stipend from the
 International Max Planck Research School (IMPRS) for Astronomy and
Astrophysics at the Universities of Bonn and Cologne.
T.~P. acknowledges support from the Combined Array for Research
in Millimeter-wave Astronomy (CARMA), which is supported by the National
Science Foundation through grant AST 05-40399.
\end{acknowledgements}

\bibliographystyle{aa} 
\bibliography{g11b}

\end{document}